\documentclass[journal,10pt]{IEEEtran}
\usepackage{amsmath,amsfonts}
\usepackage{threeparttable}
\usepackage{graphicx}
\usepackage{siunitx}
\usepackage{multirow}
\usepackage{makecell}
\usepackage{array}
\usepackage{amsmath}
\usepackage{amssymb}
\usepackage{epstopdf}
\usepackage{float}
\usepackage{amssymb}
\usepackage{nomencl}
\usepackage{enumerate}
\usepackage[linesnumbered,ruled,vlined]{algorithm2e}
\usepackage{subfigure}
\usepackage{bm}
\usepackage{amsthm}
\usepackage{siunitx}

\usepackage{multirow}
\usepackage{array}
\usepackage[caption=false,font=normalsize,labelfont=sf,textfont=sf]{subfig}
\usepackage{textcomp}
\usepackage{stfloats}
\usepackage{url}
\usepackage{xcolor} 
\usepackage{verbatim}
\usepackage{graphicx}
\usepackage{cite}
\usepackage{booktabs}
\newcommand{\RNum}[1]{\uppercase\expandafter{\romannumeral #1\relax}}
\usepackage[english]{babel}
\newtheorem{definition}{Definition}
\newtheorem{theorem}{{Theorem}}

\begin{document}

\title{A Nested Graph Reinforcement Learning-based Decision-making Strategy for Eco-platooning}


\author{
Xin Gao,~\IEEEmembership{Graduate Student Member,~IEEE}, Xueyuan Li*, Hao Liu, Ao Li,  Zhaoyang Ma,~\IEEEmembership{Graduate Student Member,~IEEE},    Zirui Li,~\IEEEmembership{Graduate Student Member,~IEEE}
\thanks{This work was supported by the Fundamental Research Funds for the Central Universities under Grant 2024CX06073.}
\thanks{
*Corresponding author: Xueyuan Li 

Xin Gao, Xueyuan Li, Hao Liu, Ao li, and Zirui Li  are with the School of Mechanical Engineering, Beijing Institute of Technology, Beijing, China. (E-mails: x.gao@bit.edu.cn; lixueyuan@bit.edu.cn; 3220210265@bit.edu.cn; 3120230295@bit.edu.cn;   z.li@bit.edu.cn.

Zhaoyang Ma is with the School of Computer and Information Technology, Beijing Jiaotong University, Beijing, China. (E-mails:zhy.ma@bjtu.edu.cn)

Xin Gao is also with the Center for AI Safety and Governance, Institute for AI, Peking University, Beijing 100871, China.

Zirui Li is also with the Chair of Trafﬁc Process Automation, ”FriedrichList” Faculty of Transport and Trafﬁc Sciences, TU Dresden, Germany.
}
}

\markboth{Journal of \LaTeX\ Class Files,~Vol.~14, No.~8, August~2021}%
{Shell \MakeLowercase{\textit{et al.}}: A Sample Article Using IEEEtran.cls for IEEE Journals}

\maketitle

\begin{abstract}
Platooning technology is renowned for its precise vehicle control, traffic flow optimization, and energy efficiency enhancement. However, in large-scale mixed platoons, vehicle heterogeneity and unpredictable traffic conditions lead to virtual bottlenecks. These bottlenecks result in reduced traffic throughput and increased energy consumption within the platoon. To address these challenges, we introduce a decision-making strategy based on nested graph reinforcement learning. This strategy improves collaborative decision-making, ensuring energy efficiency and alleviating  congestion. We propose a theory of nested traffic graph representation that maps dynamic interactions between vehicles and platoons in non-Euclidean spaces. By  incorporating spatio-temporal weighted graph into a multi-head attention mechanism, we further enhance the model's capacity to process both local and global data. Additionally, we have developed a nested graph reinforcement learning framework to enhance the self-iterative learning capabilities of platooning. Using the I-24 dataset, we designed and conducted comparative algorithm experiments, generalizability testing, and permeability ablation experiments, thereby validating the proposed strategy's effectiveness. Compared to the baseline, our strategy increases throughput by 10\% and decreases energy use by 9\%. Specifically, increasing the penetration rate of CAVs significantly enhances traffic throughput, though it also increases energy consumption.
\end{abstract}


\IEEEpeerreviewmaketitle

\section{Introduction}
\label{sec:introduction}

\IEEEPARstart{A}{s} vehicle technology evolves, especially in connectivity and automation, road traffic systems are increasingly transitioning into a hybrid intelligent transportation system \cite{Baruch,8744265}. This system integrates human-driven vehicles (HVs), autonomous vehicles (AVs), and connected and automated vehicles (CAVs), marking a pivotal shift towards more sustainable energy utilization in the transportation sector \cite{schwarting2018planning,wu2023human}. Through precise coordination and control, platooning technology offers a promising approach for optimizing traffic flow and enhancing energy efficiency \cite{10423410}.

In the realm of CAVs, numerous researchers significantly advanced energy efficiency through innovative platooning technologies \cite{9906438,han2022strategic,9585638,9410239}. Researchers such as Li et al. \cite{9410239} have developed algorithms to optimize communication strategies that manage platoon dynamics through parameter sharing and enhanced protocols, significantly improving rewards and fuel economy. Prathiba et al.  \cite{9585638}  combined deep reinforcement learning (DRL) with genetic algorithms  to develop an innovative platooning technique, enhancing cooperative driving efficiency and effectively reducing traffic congestion and fuel consumption. Additionally, Lian et al. \cite{10273625} introduced a predictive multi-agent actor-critic control framework that utilizes environmental dynamics to increase energy efficiency \cite{10381514}. Lu et al. refined the model for calculating communication energy consumption by integrating vehicular mobility factors, considering the dynamic nature of communication distances, and addressing multi-objective optimization challenges with DRL \cite{lu2022cooperative}.

\begin{figure*}[bt!]
    \centering
    \includegraphics[width=1\linewidth]{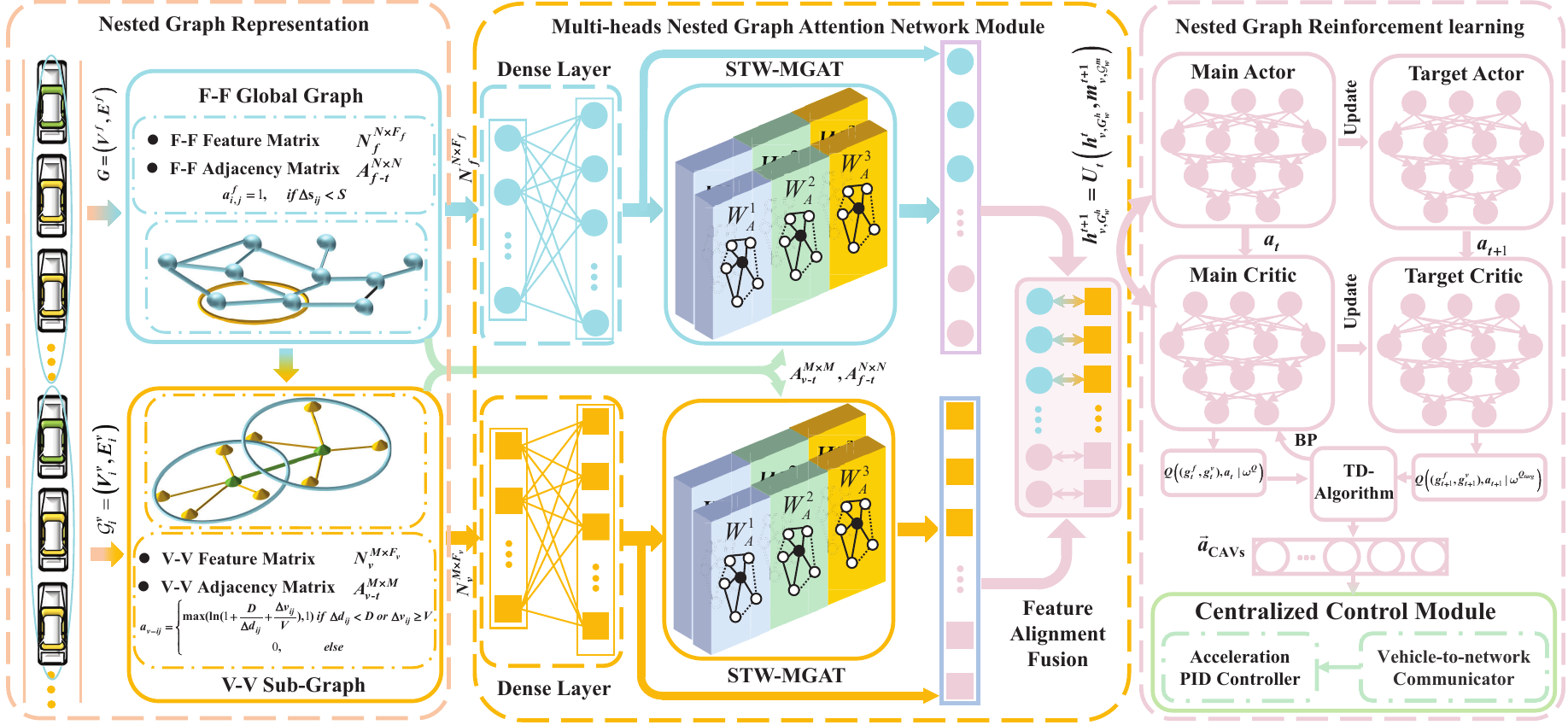}
    \caption{ The proposed framework is depicted in the schematic diagram, which employs a nested graph reinforcement learning-based decision-making model. Initially, mixed platooning scenarios are modeled using our developed nested traffic graph theory. This model is augmented by integrating the nested graph representation with a multi-head attention mechanism, resulting in a multi-head nested graph attention network module. Designed to capture the intricate spatio-temporal dependencies inherent in mixed platooning, this network aims for effective analysis. The decision-making process culminates with the generation of acceleration actions for CAVs through nested graph reinforcement learning. These actions are managed by a centralized control module, which orchestrates the CAVs within the platoon, thereby facilitating decisions that enhance safety, efficiency, comfort, and energy conservation.}
    \label{fig:frame}
\end{figure*}
However, despite significant advancements in platooning technology that have enhanced energy efficiency, its application in real-world traffic systems still confronts challenges stemming from virtual bottlenecks induced by diverse levels of vehicle intelligence \cite{10233027,8368196}. These virtual bottlenecks are often exacerbated by minor disturbances caused by HVs and less intelligent AVs, such as abrupt speed changes by the leading vehicle. These disruptions lead to errors in maintaining proper following distances, which propagate and result in repetitive stop-and-go traffic patterns \cite{8818624}. In virtual bottlenecks, repeated acceleration and deceleration cycles significantly dissipate energy across the entire platoon. Addressing virtual bottlenecks requires surmounting several hurdles \cite{li2021interactive}. Firstly, mixed platooning entails significant heterogeneity in vehicle intelligence, communication capabilities, and driving behaviors, which can result in inconsistent vehicle responses, thus compromising the platoon's efficiency and safety \cite{9737678}. Secondly, the unpredictability of HVs in this mixed traffic environment can undermine the stability and effectiveness of platoon operations  \cite{10509567}.

Although previous platooning optimization approaches have been effective in implementing cooperative control strategies for small-scale platoons \cite{9969583,li2021interactive,XIAO2024102250,9851435}, current research encounters several notable challenges: 1) Research on large-scale mixed platooning is scarce, with most studies concentrating on small, homogeneous CAV platoons, thus limiting their effectiveness in reducing traffic congestion and enhancing energy efficiency; 2) Existing research often neglects the spatial and temporal interactions between vehicles and within platoons, reducing the accuracy of essential characteristic information for effective platooning; 3)  Automation and connectivity incur costs, and it remains uncertain whether the energy used by additional equipment can be offset by smoother platooning strategies.

In this study, we propose a multi-objective optimization decision-making strategy based on nested graph reinforcement learning, designed to address the challenges of vehicle heterogeneity in large-scale mixed platooning. Utilizing a nested traffic graph representation, we capture the dynamic spatiotemporal interactions among vehicles and platoons within non-Euclidean spaces. A multi-heads nested graph attention network module is designed to effectively extract high-quality structural and feature information from both vehicle nodes and platoon subgraphs. Furthermore, we established a nested graphical Markov decision process to facilitate training our strategy, focusing on enhancing safety, task efficiency, comfort, and energy conservation. Finally, we validated our approach through a platooning simulation experiment conducted on  segment of I-24 near Nashville, Tennessee. The overall proposed framework  is illustrated in Fig.~\ref{fig:frame}. The principal contributions of this paper are outlined as follows.

\begin{enumerate}
\item A decision-making framework based on nested traffic graph theory is developed to address the challenges posed by vehicular heterogeneity in mixed platoons, which often leads to virtual bottlenecks. This framework comprises nested traffic graph representation, a nested graph Markov decision process (NG-MDP), a multi-objective dense reward model, and a multi-head nested graph attention network.
\item The nested graph representation technique captures dynamic spatio-temporal interactions across various levels within non-Euclidean spaces, identifying and processing non-homogeneous cyclic graph structures and enhancing node feature information accuracy by leveraging specific structural characteristics.
\item The dynamic weights adjacency matrix merges node features with spatio-temporal information from traffic scenarios, effectively representing the impact of spatio-temporal changes on vehicle interactions. Coupled with a multi-head graph attention mechanism, it significantly enhances the model’s capacity to process both local and global information.
\item Through an extensive series of simulation experiments, we evaluated crucial factors such as traffic flow, vehicle speed distribution, energy efficiency, and congestion conditions. Importantly, our findings elucidate the scientific principles underlying the balance between managing overall energy consumption and enhancing traffic mobility in large-scale mixed platoons, facilitated by advancements in connectivity and automation.
\end{enumerate}

The remainder of this article is organized as follows.  The proposed Nested graph theory for traffic representation is illustrated in section \RNum{2}. Section \RNum{3} introduces the methodology, including a nested graph reinforcement learning framework. Section \RNum{4} explains the   setting of   experiments. Section \RNum{5}   analyzes the simulation results. Conclusions are drawn in section \RNum{6}.

\begin{figure}[!t]
\centerline{\includegraphics[width=\columnwidth]{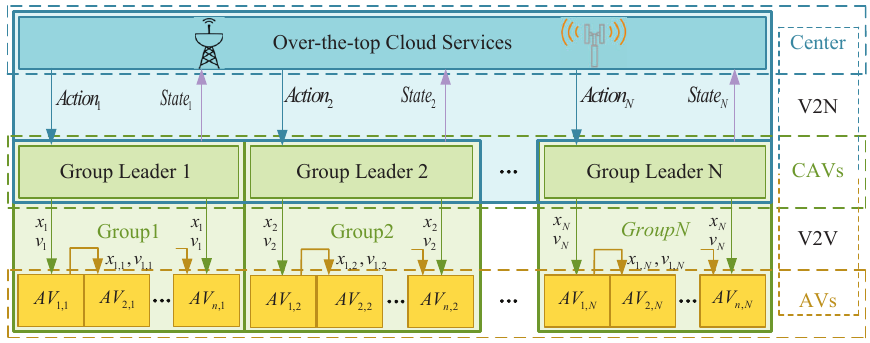}}
\caption{The hierarchical platoon architecture.}
\label{fig2}
\end{figure}

\section{Nested graph theory for traffic representation}
\label{sec:theory}
This section delves into the application of nested graph theory for managing large-scale autonomous vehicle platoons. By establishing a hierarchical platooning architecture and utilizing nested graph representations to characterize the traffic environment, this approach adeptly captures dynamic spatiotemporal interactions between vehicles and platoons. It further extends the Nested Graphical Markov Decision Process.

\begin{figure}[!t]
\centerline{\includegraphics[width=\columnwidth]{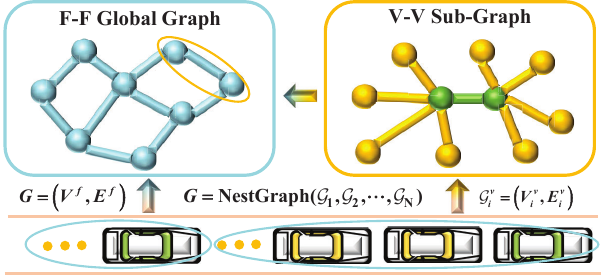}}
\caption{A schematic diagram of a nested traffic graph.}
\label{fig3}
\end{figure}
\subsection{Hierarchical platooning architecture}
In response to the demands for large-scale autonomous vehicle platooning, a hierarchical platoon architecture has been established, as illustrated in Fig. \ref{fig2}.  This architecture is conceptualized as a vehicular edge network and comprises a unidirectional roadway, a single cloud server, \( R \) roadside units (RSUs) distributed along the roadway, and \( N \) groups. In each group, \( N \) AVs are led by a CAV. The operational logic of this system is outlined as follows.
\begin{enumerate}
\item The leader of each group, a CAV, utilizes Vehicle-to-Network (V2N) communication for transmitting data about its group to the cloud and receiving operational directives from the cloud.
\item Within each group, the AVs, denoted as \( \{ AV_{1,j}, AV_{2,j}, \ldots, AV_{n,j} \} \), and the \( CAV_j \) utilize vehicle-to-vehicle (V2V) communication protocols. The number of AVs, \( n \), within each group is determined by the minimum transmission range required for effective relay among AVs. These AVs follow the CAV using the intelligent driver model (IDM) \cite{treiber2000congested}.
\item A binary variable \( x_{i,e} \), where \( e \in [0, R + 1] \) and \( x_{i,e} \in \{ 0, 1 \} \), is defined to indicate whether the computational task of vehicle \( i \) is assigned to server \( e \) for execution. If \( x_{i,0} = 1 \), the vehicle \( i \) performs the computation locally; if \( x_{i,R + 1} = 1 \), the vehicle \( i \)'s task is offloaded to the cloud server; otherwise, the computation is offloaded to the corresponding vehicular edge computing (VEC) server.
\end{enumerate}

\subsection{Nested graph theory for traffic representation}\label{section3.2}


In dynamic platooning scenarios, the interactions among vehicles and between platoons exhibit spatiotemporal dynamics and nested complexity. In mixed platooning environments, it is essential to address not only vehicle-to-vehicle (V-V) interactions but also the comprehensive interactions between formations (F-F). As vehicles alter their positions in space, the nature and intensity of these interactions vary, demonstrating rich spatiotemporal dynamics. To model these interactions, nested and weighted graphs are utilized.
\subsubsection{Definition}
\begin{definition}[Nested traffic graph]
Define the traffic environment at each time \(t\) as a nested graph \(G = \text{NestGraph}(\mathcal{G}_1, \mathcal{G}_2, \ldots, \mathcal{G}_N)\), where \(\text{NestedGraph}(\cdot)\) denotes the function used to merge nested graphs. \(\mathcal{G}_N\) is the \(N\)-th subgraph, representing the total number of formation groups. Each \(i\)-th V-V subgraph, \(\mathcal{G}_i^v = (V_i^v, E_i^v)\), where \(i \in \{1, 2, \ldots, N\}\), consists of nodes \(V_i^v\) with each vehicle represented as a node \(u \in V_i^v\). Information interaction between vehicle \(a\) and vehicle \(b\) is captured as \( (u_a, u_b) \in E_i^v \). The global F-F graph, \(G = (V^f, E^f)\), includes nodes \(V^f\) where each platoon is a node \(u^f \in V^f\), and interactions between platoon \(i\) and platoon \(j\) are represented as \( (u_i^f, u_j^f) \in E^f \). The graph over time features an F-F node feature matrix \(N_{f-t}^{N \times F_f}\) and an adjacency matrix \(A_{f-t}^{N \times N}\), with \(F_f\) representing the total features per platoon. All V-V subgraphs are interconnected, depicted through the vehicle subgraph node feature matrix \(N_{v-t}^{M \times F_v}\) and the vehicle subgraph adjacency matrix \(A_{v-t}^{M \times M}\), where \(F_v\) is the feature count per vehicle and \(M\) is the total number of vehicles. A schematic diagram of a nested traffic graph is illustrated in Fig. \ref{fig3}.
\end{definition}

\begin{figure}[!t]
\centerline{\includegraphics[width=\columnwidth]{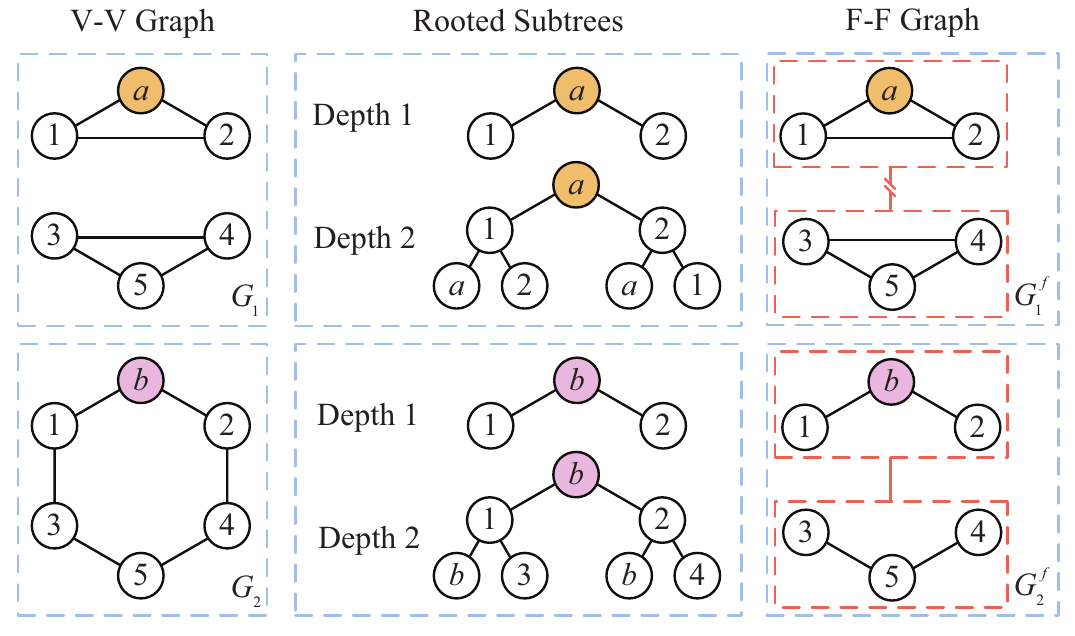}}
\caption{Distinction of non-isomorphic cyclic traffic graphs.}
\label{fig:root}
\end{figure}
\subsubsection{Non-homogeneous cyclic graph}

As depicted in Fig.~\ref{fig:root}, the diagrams $G_1$ and $G_2$ exemplify typical formations observed during convoy maneuvers. Notably, $G_1$ illustrates sparse spatiotemporal interactions between the vehicle pairs 1, 2 and 3, 4, while $G_2$ demonstrates a closer spatiotemporal arrangement among these vehicles, thereby enhancing inter-vehicle communication. Utilizing these two distinct, non-isomorphic cyclic vehicle subgraphs as case studies, conventional Graph Neural Networks (GNNs) at a root level depth of one exhibit limitations in differentiation. Regardless of iteration count, both graphs persistently display identical representations due to the homogeneity of all nodes within any root subtree at height one. However, from an integrative standpoint, these configurations are markedly divergent: $G_1$ is constituted by two separate triangles, whereas $G_2$ forms a hexagonal structure. Consequently, traditional GNNs can only differentiate between $G_1$ and $G_2$ through progressive increases in network depth, which in turn precipitates a surge in computational demands.

In contrast, a nested graph representation strategy is employed, where nodes a, 1, 2, 3, 4, 5, b, 1, and 2 are treated as discrete sub-formations. Drawing upon the structures of $G_1$ and $G_2$, two novel formation graphs, $G_1^f$ and $G_2^f$, are delineated, as indicated by the red dashed lines on the right side of Fig.~\ref{fig:root}. These lines signify the absence of informational exchanges between sub-formations not encompassed within the same RSU range. Analytical results reveal that $G_1^f$ encapsulates a closed triangular configuration, while $G_2^f$ exhibits an open triangular framework. This methodology substantially augments the graph's capacity to distinguish edge attributes and more effectively captures the overarching structural nuances, thus surmounting the inherent constraints of traditional GNNs in managing complex graph architectures. It adeptly addresses the prevalent challenge of diminished spatiotemporal interaction modeling accuracy in large-scale heterogeneous vehicle formations.

\subsubsection{ Nested graph message passing}
To further enhance the representation of structural and feature information within vehicle nodes and platoon subgraphs, we propose the integration of nested graph message passing and update functions. The process initiates with message passing between the global graph \( G_\omega^h \) and the corresponding subgraph \( \mathcal{G}_\omega^m \) at node \( \omega \). The equations governing these interactions are given as
\begin{equation}
h_{v, G_{\omega}^h}^{t+1} = U_t\left(h_{v, G_{\omega}^h}^t, m_{v, \mathcal{G}_{\omega}^m}^{t+1}\right),
\end{equation}
\begin{equation}
m_{v, \mathcal{G}_{\omega}^m}^{t+1} = \sum_{u \in N(v| \mathcal{G}_{\omega}^m)} M_t\left(h_{v, \mathcal{G}_{\omega}^m}^t, h_{u, \mathcal{G}_{\omega}^m}^t, e_{vu}\right),
\end{equation}
where \( U_t(\cdot) \) and \( M_t(\cdot) \) represent the update and message functions of the GNN at time step \( t \), and \( N(v| \mathcal{G}_{\omega}^m) \) denotes the set of neighboring nodes within the subgraph \( \mathcal{G}_{\omega}^m \).

\subsubsection{Extended theorem}
Post-aggregation, node representations are summarized through a pooling layer to support graph-level tasks. However, representing diverse substructures in cyclic graphs remains a challenge. Mixed platooning scenarios, characterized by their spatiotemporal dynamics and multiple dynamic cyclic subgraphs, complicate the task further. Therefore, traditional GNNs often fail to accurately distinguish between non-isomorphic cyclic graphs. In contrast, NGNNs offer improved discrimination at both the node and subgraph levels through their enhanced structural sensitivity.

Theorem 1 formalizes the capability of NGNNs to discern non-isomorphic cyclic graphs \cite{zhang2021nested}.
\begin{theorem}
Given two nested graphs \( G_1 \) and \( G_2 \), if at least one level of subgraphs exhibits distinct structural entropy in their representations, then the total entropy of these nested graphs also differs, i.e., \( H(\text{NG}(G_1)) \ne H(\text{NG}(G_2)) \). Each subgraph \( \mathcal{G}_i \) is characterized by a Laplacian matrix \( L_i = D_i - A_i \), where \( D_i \) is the degree matrix and \( A_i \) is the adjacency matrix. The eigenvalues of \( L_i \) are denoted as \( \lambda_{i1}, \lambda_{i2}, \ldots, \lambda_{ik_i} \), leading to a spectral entropy defined as
\begin{equation}
H(\mathcal{G}_i) = -\sum_{j=1}^{k_i} \frac{\lambda_{ij}}{\sum_{l=1}^{k_i} \lambda_{il}} \log \left(\frac{\lambda_{ij}}{\sum_{l=1}^{k_i} \lambda_{il}}\right).
\end{equation}

The overall entropy of the nested graph is calculated as a weighted average of the entropies of its subgraphs
\begin{equation}
H(\text{NG}(G)) = \sum_{i=1}^n w_i H(\mathcal{G}_i).
\end{equation}

Differences in entropy at any subgraph level imply discrepancies in the total entropy, thereby establishing the non-isomorphism of \( G_1 \) and \( G_2 \).
\end{theorem}

By leveraging detailed feature information from subgraphs, NGNNs significantly improve the quality of node representations, particularly through specific structural features. Additionally, the transmission of information in mixed platooning scenarios through nested and underlying graph structures has been quantified, resulting in the following theorem.

\begin{theorem}
A nested graph comprising multiple sub-vehicle formations exhibits an information intensity that is at least equal to the sum of the information intensities of its constituent vehicle subgraphs. The interaction intensity between any two nodes \( u, v \) in a subgraph \( \mathcal{G}_i \) is defined as \( I(u, v) \). The total information intensity for a subgraph \( \mathcal{G}_i \) is expressed as
\begin{equation}
S(\mathcal{G}_i) = \sum_{u, v \in \mathcal{G}_i} I(u, v).
\end{equation}

Information interactions between different platoons can be quantified as
\begin{equation}
S_{\text{inter}}(\text{NG}(G)) = \sum_{i \ne j} \beta_{ij} \sum_{u in \mathcal{G}_i, v in \mathcal{G}_j} I(u, v) \ge 0,
\end{equation}
where \( \beta_{ij} \) represents the interaction intensity between platoons \( i \) and \( j \). The cumulative information intensity of the nested graph, combining both internal and interaction components, is formulated as
\begin{equation}
S(\text{NG}(G)) = \sum_i S(\mathcal{G}_i) + S_{\text{inter}}(\text{NG}(G)) \ge \sum_i S(\mathcal{G}_i).
\end{equation}
\end{theorem}

\subsection{Nested graphical markov decision process}\label{section:NGMDP}

The introduction of mixed platooning presents unique challenges in coordinating decisions among heterogeneous CAVs. To address this complexity, we propose the Nested Graphical Markov Decision Process (NGMDP), an extension of the conventional Markov Decision Process (MDP) framework \cite{van2012reinforcement}. This adaptation is designed to handle the intricate dynamics of mixed platooning scenarios effectively.

\begin{figure}[!t]
\centerline{\includegraphics[width=\columnwidth]{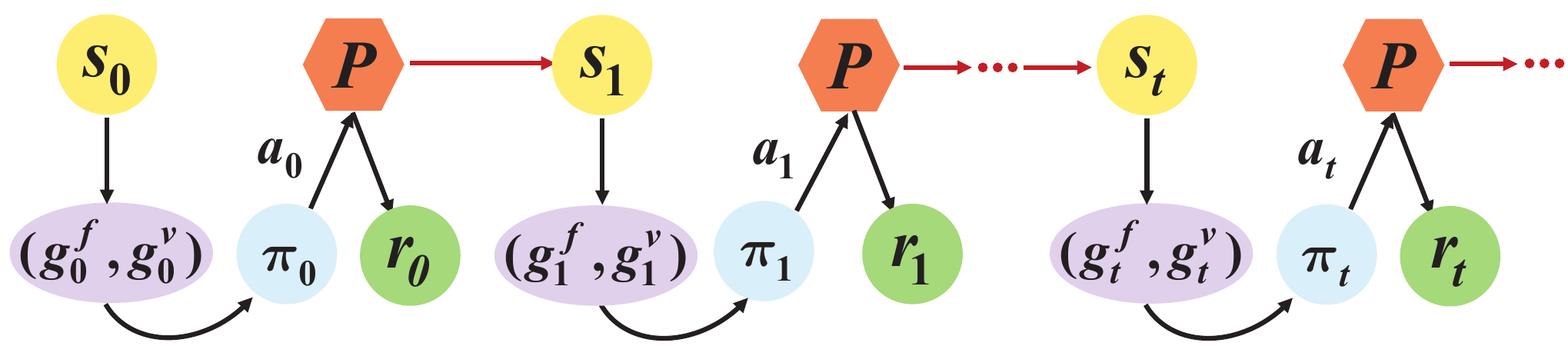}}
\caption{Nested graphical markov decision process.}
\label{fig:markov}
\end{figure}

\begin{definition}[Nested graphical markov decision process]
The NGMDP is defined by a tuple \(\mathcal{N} = (N, M, \mathcal{S}, \mathcal{G}, \mathcal{A}, \mathcal{P}, \mathcal{R})\). The objective is to develop a joint policy \(\pi\) that maximizes the global value function \(Q^\pi((g_f, g_v), a)\).
\begin{itemize}
    \item \(N\) represents the number of platoons, denoted as \(\{F_1, F_2, \ldots, F_N\}\).
    \item \(M\) is the total number of vehicles, represented as \(\{V_1, V_2, \ldots, V_M\}\).
    \item \(\mathcal{S}\) denotes the state information of all vehicles within the environment, expressed as \((s_1, s_2, \ldots, s_M) \in \mathcal{S}\).
    \item \(\mathcal{G} = \{\mathcal{G}_f, \mathcal{G}_v\}\) signifies the nested graph representation of inter-platoon and inter-vehicle information derived from \(\mathcal{S}\), with elements \((g_1^{\mathcal{T}}, g_2^{\mathcal{T}}, \ldots, g_n^{\mathcal{T}}) \in \mathcal{G}_{\mathcal{T}}, \mathcal{T} \in \{f, v\}\).
    \item \(\mathcal{A}\) comprises all potential actions initiated by platoon leaders, with \((A_1, A_2, \ldots, A_N) \in \mathcal{A}\).
    \item \(\mathcal{P}\) represents the probabilities of transitioning from a current graphical state \((g_f, g_v)\) to a subsequent graphical state \((g_f', g_v')\), denoted as \(P((g_f', g_v') | (g_f, g_v), a) \in \mathcal{P}\).
    \item \(\mathcal{R}\) calculates rewards based on actions taken in a graphical state, defined as \(R((g_f, g_v), a) = \sum_{i=1}^N r((g_f^i, g_v^i), a_i)\), where each \((g_f^i, g_v^i)\) and \(a_i\) represent the local state and action for the \(i\)-th CAV, respectively.
\end{itemize}

\end{definition}

The operation of the NGMDP is illustrated in Fig. \ref{fig:markov}. At each timestep \(t\), the current environmental state \(s_t\) is processed through nested graph representation theory, producing the F-F and V-V graphical states \((g_t^f,g_t^v)\). These states inform the policy network \(\pi_t\), which generates an appropriate action \(a_t\). This action, in conjunction with the graphical state, influences state updates via the transition probability \(P\). Rewards are then computed based on the current environment using the reward function \(r_t\), which in turn informs subsequent updates to the policy network \(\pi_t\).

\section{Methodology}

Detailed herein is a multi-objective optimization decision strategy based on nested graph reinforcement learning, which optimizes interactions and decision-making processes within the platoon by defining state and action spaces. Moreover, the enhancement of the model’s processing capabilities through spatiotemporal dynamic adjacency matrices and multi-headed nested graph attention network modules is thoroughly discussed.

\subsection{Problem formulation}
\subsubsection{State space} 

The state space builds upon the graphical observed states previously outlined in Section 3.1, \(\mathcal{G} = \{\mathcal{G}_f, \mathcal{G}_v\}\). We detail the state information for platoons and individual vehicles as follows

For the \(i\)-th platoon, the state \(s_i^f\) captures collective dynamics and their impact on traffic flow, defined by the equation
\begin{equation}
\mathbf{s_i^f} = [\mathbf{V_i}, \mathbf{\Delta X_i}, \mathbf{\bar{V}_i}, \mathbf{\overline{a}_i}],
\end{equation}
where \(\mathbf{V_i}\) is the speed of the platoon's lead CAV. \(\mathbf{\Delta X_i}\) denotes the relative distance to the preceding vehicle. \(\mathbf{\bar{V}_i}\) is the average speed of the platoon. \(\mathbf{\overline{a}_i}\) represents the average acceleration within the platoon.

The state for the \(j\)-th vehicle, \(\mathbf{s_j^v}\), includes detailed characteristics, described by
\begin{equation}
\mathbf{s_j^v} = [\mathbf{X_j}, \mathbf{V_j}, \mathbf{I_j}, \mathbf{\Delta X_j}, \mathbf{\Delta V_j}, \mathbf{a_j}],
\end{equation}
where \(\mathbf{X_j}\) is the longitudinal position. \(\mathbf{V_j}\) represents the vehicle's speed. \(\mathbf{I_j}\) indicates the vehicle type. \(\mathbf{\Delta X_j}\) is the relative distance to the vehicle ahead. \(\mathbf{\Delta V_j}\) is the relative speed with respect to the preceding vehicle.\(\mathbf{a_j}\) denotes the vehicle's acceleration.

\subsubsection{Action space} 

The action space for CAVs within the platoons is defined to facilitate simulation and optimization of decision-making behaviors
\begin{equation}
\mathbf{\mathcal{A}} = \left\{\mathbf{ a} \in \mathbf{\mathbb{R}^N }: -4.5 \leq a_i \leq 4.5, \forall i \in \{1, 2, \ldots, N\} \right\},
\end{equation}
where \(a_i\) specifies the acceleration for the \(i\)-th CAV, constrained within a continuous interval.

\subsection{ Spatio-temporal dynamic  adjacency matrix}\label{SWG}

In vehicular traffic scenarios, weighted graphs are crucial for accurately modeling the spatiotemporal dynamics of vehicle interactions, as opposed to uniform-weight graphs which offer a more simplified overview. The edge weights in these graphs represent the functional dependencies between nodes, encapsulating the dynamic interactions over time and space. This detailed representation is achieved through spatiotemporal weighted adjacency matrices, which are tailored for both global and subgraph levels within nested traffic graph frameworks.

The formation of the F-F adjacency matrix \( A_{f-t}^{N \times N} \) is predicated on the following assumptions:
\begin{enumerate}[(a)]
    \item Within a given RSU communication range, all platoons are assumed to be capable of sharing information, indicated by \( a_{i,j}^f = 1 \) for \( i \ne j \).
    \item Self-information sharing within platoons is represented by \( a_{i,i}^f = 1 \).
\end{enumerate}

Similarly, the V-V subgraph adjacency matrix \( A_{v-t}^{M \times M} \) is computed based on these assumptions:
\begin{enumerate}[(a)]
    \item All CAVs within the same RSU range can share information, denoted by \( a_{i,j}^v = 1 \).
    \item Direct information sharing between AVs is not possible.
    \item CAVs are capable of sharing information regarding AVs within their communication reach.
    \item Self-information sharing among vehicles is allowed and is denoted by \( a_{i,i}^v = 1 \).
\end{enumerate}

To model the interactions between CAVs and AVs more precisely, we introduce a spatiotemporal weighted adjacency matrix that accounts for the relative distance and speed between vehicles. The weight \( a_{i,j}^v \) is defined by the following relationship:
\begin{equation}
a_{i, j}^v=\left\{\begin{array}{l}
\max \left(\ln \left(1+\frac{D_{\max }}{\Delta d_{i, j}}+\frac{\left|\Delta v_{i, j}\right|}{V_{\max }}\right), 1\right) \\
\quad \quad \text { if } \Delta d_{i, j}<D_{\max } \text { and }\left|\Delta v_{i, j}\right| \geq V_{\max } \\
0, \quad \quad \quad \quad \quad  \quad \quad \quad \quad \quad \quad \quad \quad\text { otherwise }
\end{array}\right.
\end{equation}
where \( D_{\text{max}} \) and \( V_{\text{max}} \) are predefined thresholds for distance and velocity, respectively. This formulation ensures that the weight approaches 0 as the relative distance \( \Delta d_{i,j} \) nears the maximum allowable distance, or the relative velocity \( \Delta v_{i,j} \) approaches the velocity limit. Conversely, the weight approaches 1 as the relative distance decreases or the relative velocity increases, enhancing the model's sensitivity to critical spatiotemporal variations.
\subsection{ Multi-Head Nested Graph Attention Network Module} \label{MGAT}
To integrate seamlessly with the nested graph representation, we propose a multi-head nested graph attention network (MNGAT) to adeptly manage feature information across various dimensions \cite{veličković2018graph}. Attention weights for the nested node feature set \(\mathbf{h} = \{h_1, h_2, \ldots, h_M\}\) are calculated as follows
\begin{equation}
\alpha_{i,j} = \frac{\exp\left(\text{LeakyReLU}\left(a^T(W h_i \oplus W h_j)\right)\right)}{\sum_{k \in \mathcal{N}_i} \exp\left(\text{LeakyReLU}\left(a^T(W h_i \oplus W h_k)\right)\right)},
\end{equation} 
where \(W\) is a learnable projection matrix, \(a^T\) is a trainable attention vector, and \( \oplus \) denotes vector concatenation. To enhance the learning capability of the model, a multi-head attention mechanism aggregates the outputs of several independent attention heads, using averaging to manage dimensionality in the final layer
\begin{equation}
h'_i = \frac{\sigma}{K} \sum_{k=1}^K \sum_{j \in \mathcal{N}_i} \left(\alpha_{i,j}^k W^k h_j\right), 
\end{equation}  
where \(\sigma\) is an activation function, \(K\) the number of attention heads, \(\alpha_{i,j}^k\) the attention weights for the \(k\)-th head, and \(W^k\) the corresponding transformation matrices.
\subsection{Reward model} 

The reward model integrates multiple objectives to promote safety, task completion, comfort, and energy efficiency. We define the reward vector as
\begin{equation}
\mathbf{R} = [\mathbf{R_{\text{safe}}}, \mathbf{R_{\text{task}}}, \mathbf{R_{\text{comfort}}}, \mathbf{R_{\text{energy}}}],
\end{equation}

\subsubsection{Safety sub-reward function}\label{safe}
The safety sub-reward incorporates metrics for braking distances and time to collision (TTC) to mitigate risks effectively
\begin{equation}
R_{\text{safe}, D_s} = \begin{cases} 
-1 & \text{if } \Delta d < D_s \\
0 & \text{if } \Delta d \geq D_s 
\end{cases} ,
\end{equation}
\begin{equation}
R_{\text{safe}, TTC} = \begin{cases} 
\max \left\{ -1, \ln \left( \frac{TTC}{T_{\text{limit}}} \right) \right\} & \text{if } 0 < TTC \leq T_{\text{limit}} \\
0 & \text{if } TTC > T_{\text{limit}}
\end{cases} ,
\end{equation}
 where \(D_s = D_1 - D_2 + d_0 = v_e t_0 + \frac{v_e^2 - v_f^2}{2a_{\max}} + d_0\), \(\Delta d\) is the distance to the vehicle ahead, \(D_s\) is the safe distance, \(T_{\text{limit}}\) is the TTC threshold set at 4 seconds \cite{ZHU2020102662}, \(D_1\) is the distance traveled from perceiving a hazard to the vehicle stopping, \(D_2\) is the distance the front vehicle travels when it suddenly brakes to a complete stop, \(d_0\) is the minimum vehicle spacing, \(v_e\) is the ego vehicle speed, \(t_0\) is the communication delay time, \(a_{\max}\) is the maximum deceleration, and \(v_f\) is the front vehicle speed.

\subsubsection{Task sub-reward function} \label{task}
Task-oriented rewards focus on maintaining a predefined following distance and matching speeds with the preceding vehicle
\begin{equation}
R_{\text{task},\Delta d} = -\tanh\left(\left|\Delta d - \Delta d_{\text{des}}\right|\right),
\end{equation}
\begin{equation}
R_{\text{task},\Delta v} = -\tanh\left(\left|v_{\text{ego}} - v_f\right|\right),
\end{equation}
where \(\Delta d_{\text{des}}\) specifies the desired following distance.
\IncMargin{1em}
\begin{algorithm} \SetKwData{Left}{left}\SetKwData{This}{this}\SetKwData{Up}{up} \SetKwFunction{Union}{Union}\SetKwFunction{FindCompress}{FindCompress} \SetKwInOut{Input}{input}\SetKwInOut{Output}{output}
	
        \Input{ Initialize replay buffer ${{\cal D}}$,training batch size ${U}$, actor network parameters $\theta $, target actor network parameters $\bar \theta  \leftarrow \theta $, Value network parameters $\omega $, target value network parameters ${\bar \omega} \leftarrow \omega$, V-V graph attention  parameters ${\vartheta ^v}$, F-F graph attention  parameters ${\vartheta ^f}$}
        
	 \BlankLine 
	 \For{episode$ \in \{ 0,1, \ldots ,M\} $}{ 
	 	\emph{Initialize the noise of exploration $\varepsilon $, receive initial observation state $s_0$, graph observation states $(g_1^v,g_1^f)$.}
            
	 	\For{t $ \in \{ 0,1, \ldots ,T\} $}{\label{forins} 
            
                  \emph{Feature alignment fusion:${F_t} = NGAT\left( {g_t^v;{\vartheta ^v}} \right)||NGAT\left( {g_t^f;{\vartheta ^f}} \right)$ }

                  \emph{Select action:${a_t} = \mu \left( {{F_t};\theta } \right) + \varepsilon $ }
                  
                  \emph{Obtain reward \(r_t\), new state $s_{t+1}$ and $(g_{t+1}^v,g_{t+1}^f)$ }

                  \emph{Add tuple $ ({s_t},{(g_{t}^v,g_{t}^f),a_t^F,r_t^F,{s_{t + 1}},(g_{t+1}^v,g_{t+1}^f))}$ into ${{\cal D}}$ }

                  \emph{Sample ${U}$ tuples from ${{\cal D}}$ }

                  \emph{Update value network: $L(w) = \frac{1}{N}\sum\limits_{i = 0}^{N - 1} {{{\left[ {Q\left( {{F_i},{a_i};w} \right) - {y_i}} \right]}^2}} $}

                  \emph{Update actor network: ${\nabla _\theta }J = \frac{1}{N}\sum\limits_{i = 0}^{N - 1} {{\nabla _a}Q\left( {{F_i},{a_i};w} \right)}  \cdot {\nabla _\theta }\mu \left( {{F_i};\theta } \right)$}

                  \emph{Update target networks $\bar w \leftarrow \tau w + (1 - \tau )\bar w$,$\bar \theta  \leftarrow \tau \theta  + (1 - \tau )\bar \theta $}

            }
            
        \emph{\textbf{end}}
 		 }
     \emph{\textbf{end}}
   
 	 	  \caption{NSTW}
 	 	  \label{algo_disjdecomp} 
 	 \end{algorithm}
 \DecMargin{1em} 
\subsubsection{comfort sub-reward function}\label{confort}
To enhance passenger comfort, we introduce a sub-reward function that addresses both the magnitude and the change rate of acceleration (jerk) \cite{8263470}. The comfort-related reward functions are given by
\begin{equation}
R_{\text{comfort},a} = \max\left(-1, -\left(\frac{a}{C_1}\right)^2\right), 
\end{equation}   
\begin{equation}
R_{\text{comfort},\text{jerk}} = \begin{cases} 
\max\left(-1, -C_2 \frac{j(t)^2}{j_{\max}^2}\right) & \text{if } |j(t)| \geq 2.94, \\
-\frac{j(t)^2}{j_{\max}^2} & \text{if } |j(t)| < 2.94.
\end{cases} 
\end{equation}    
where \(C_1\) and \(C_2\) are tuning constants, \(j(t)\) represents the jerk at time \(t\), and \(j_{\max}\) is the maximum allowable jerk, set at 90 m/s\(^3\).

\subsubsection{ Energy sub-reward function}\label{energy}
The energy sub-reward functions target reductions in communication and computational energy expenditures \cite{qu2022automation}, thereby contributing to an overall energy-saving strategy in intelligent transportation systems \cite{10259658}. The specific functions are outlined as follows
\begin{equation}
R_{\text{energy},\text{battery}} = -\sum_{i=1}^M \left(P_{i,\text{bat,chem}} \cdot t_s\right), 
\end{equation} 
\begin{equation}
R_{\text{energy},\text{comm}}^{\text{VEC}} = -\sum_{i=1}^N E_i^{\text{comm}}, 
\end{equation}
\begin{equation}
R_{\text{energy},\text{mig}}^{\text{AVs}} = -\sum_{i=1}^N \sum_{f=1}^{N_{i,f}} E_{i,i_f}^{\text{mig}},
\end{equation}  
\begin{equation}
R_{\text{energy},\text{cal}} = -\sum_{i=1}^N E_i^{\text{local}},
\end{equation}  
where \(P_{i,\text{bat,chem}}\) represents the chemical battery power consumption for vehicle \(i\) covering motor and auxiliary systems, \(t_s\) is the time step, \(E_i^{\text{comm}}\) measures the energy used in communications with the VEC, \(N_{i,f}\) is the count of following vehicles, \(E_{i,i_f}^{\text{mig}}\) quantifies the communications migration energy between vehicle \(i\) and each follower \(i_f\), and \(E_i^{\text{local}}\) accounts for the energy expended in local calculations.

\subsection{Integration}
NSTW, through the integration of the above mechanisms, leverages nested graph representation theory, and is described as Algorithm 1. Given the extensive body of research, this paper omits discussion of  DRL algorithms.

\section{Experimental Settings} 
To validate the proposed algorithm and its energy-saving effects in networked applications, we have designed experiments for algorithm comparison, penetration rate comparison, and generalization testing.

\subsection{Simulation settings}
To validate our proposed framework, we utilized the FLOW \cite{Wu_2022,koch2021accurate} traffic simulation platform combined with Python to establish an experimental environment designed to enhance platooning system capacity. The scenario is characterized by significant speed fluctuations in the trajectory-leading vehicle (TL) due to external influences, requiring the entire mixed platoon to smoothly adapt to a dynamically changing velocity trajectory. In the experiment, 201 vehicles were simulated, with the TL's trajectory based on extreme scenarios from I-24 highway data \cite{9811912}, incorporating complex maneuvers such as emergency braking, low-speed travel, and rapid acceleration, as depicted in Fig.~\ref{fig:traject}. The platoon configuration consisted of one TL, 10 CAVs, and 190 AVs, organized into 10 groups. Detailed scenario parameters and hyperparameters are documented in Table \ref{tab:table:para}.
\begin{table}[!t]
\caption{The hyperparameters  setting.\label{tab:table:para}}
\newcolumntype{C}{>{\centering\arraybackslash}X}
\centering
\begin{tabular}{cc}
\toprule
\textbf{Parameter} & \textbf{Value}   \\
\midrule

 Time step  & 0.1 ${\rm{s}}$\\
 Total time steps &\num{9e5}\\
Minibatch size & 64 \\
 Exploration time step&  \num{6e5}\\
 Starting greedy rate & 0.5\\
 Ending greedy rate &  \num{7.5e-3}\\
  Learning rate & \num{7.5e-3}\\
Updating rate  & \num{7.5e-2}\\
Length of road &\num{2.5e4} m \\
Number of vehicles  & 201\\
Initial vehicle spacing&40 m\\
 Maximum speed & 40 m/s\\
\hline
\end{tabular}
\end{table}
\subsection{Experimental categories}
\begin{figure*}[bt!]
    \centering
    \includegraphics[width=1\linewidth]{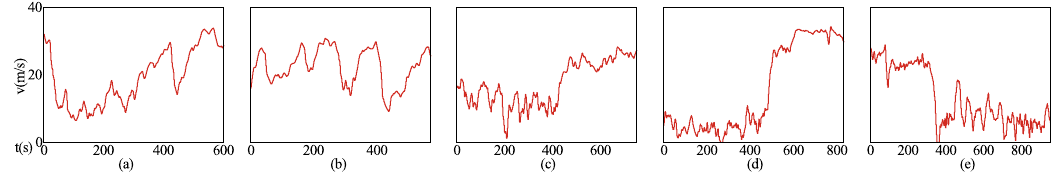}
    \caption{  Trajectories of lead vehicle  based on I-24 highway data: (a) Integrated trajectory, (b) High-speed, (c) Low-speed, (d) Rapid acceleration, (e) Emergency braking. }
    \label{fig:traject}
\end{figure*}
Experiments designed to compare algorithms, evaluate penetration rates, and test generalizability have been developed to validate the effectiveness of the proposed framework.
\begin{enumerate}[(a)]
\item Comparative algorithm experiments: These experiments involve detailed comparisons between the results from the training and testing phases. They incorporate analyses of spatiotemporal trajectories and performance metrics to provide an in-depth evaluation of formation performance across various algorithms. Furthermore, a comprehensive analysis is conducted to determine whether the optimization strategies can mitigate the increased energy consumption associated with the integration of connected and autonomous driving features. The trajectory employed by the lead vehicle is depicted in Fig.~\ref{fig:traject}(a).
\item Generalizability testing experiments: These tests assess the NSTW algorithm's ability to adapt to speed trajectories that were not part of its training set. The focus is on the algorithm’s capacity to quickly and accurately adjust to new trajectories, thereby maintaining stability and efficiency beyond its initial training parameters. The testing trajectory employed by the lead vehicle is depicted in Fig.~\ref{fig:traject}(b)-(d).

\item Permeability ablation experiments: These experiments deeply examine the effects of varying penetration rates of CAVs on traffic flow. The focus is on assessing the NSTW algorithm's overall impact on traffic stability, efficiency, and energy consumption under different CAV penetration scenarios. The trajectory employed by the lead vehicle is depicted in Fig.~\ref{fig:traject}(a).
\end{enumerate}

\subsection{Description of algorithms}
The objective of the algorithm comparison simulation experiments is to elucidate the specific impacts of various algorithmic components on overall performance. This facilitates a deeper understanding of the mechanisms and contributions of different algorithms in enhancing safety, efficiency, comfort, and energy efficiency. The five comparative algorithms under review are detailed in Table \ref{tab:table1}.
\begin{enumerate}[(a)]
\item IDM: This strategy replicates human driving behaviors and is utilized as a benchmark for evaluating the performance of competing algorithms.
\item DDPG: This model employs a deep reinforcement learning framework and does not incorporate the network topologies among vehicles, focusing instead on the exploration of vehicle control strategies in environments devoid of graphical data.
\item MGAT: An enhancement of DDPG, this approach integrates an unweighted multi-head graph attention mechanism to more intricately capture the interactions and dependencies among vehicles, thereby improving the learning efficiency of the strategy and the quality of decision-making.
\item STW: Augmenting MGAT, this strategy introduces spatial-temporal weighting as discussed in Section \ref{SWG}, enabling the algorithm to modulate attention weights based on temporal and spatial factors during vehicle interactions.
\item NSTW: Employing the nested graph attention mechanism outlined in Section \ref{MGAT}, this strategy seeks to capture more complex dependencies by simulating interactions among vehicles across multiple levels, thus significantly optimizing the decision-making process. 
\end{enumerate}

\begin{table}[!t]
\caption{Comparison of    algorithms.\label{tab:table1}}
\newcolumntype{C}{>{\centering\arraybackslash}X}
\centering
\begin{tabular}{ccccc}
\toprule
\textbf{Algorithm} & \textbf{DDPG} \cite{lillicrap2015continuous}   & \textbf{MGAT}  & \textbf{STW Graph} &  \textbf{Nested Graph}  \\
\midrule
IDM &$\times$ &$\times$ &$\times$ &$\times$\\
DDPG &\textbf{$\surd$}& $\times$&$\times$ &$\times$ \\
MGAT &\textbf{$\surd$}&\textbf{$\surd$} &$\times$ &$\times$\\
STW &\textbf{$\surd$}&\textbf{$\surd$} &\textbf{$\surd$} & $\times$\\
NSTW &\textbf{$\surd$}&\textbf{$\surd$} &\textbf{$\surd$} & \textbf{$\surd$}\\
\hline
\end{tabular}
\end{table}
\begin{figure}[!t]
\centerline{\includegraphics[width=\columnwidth]{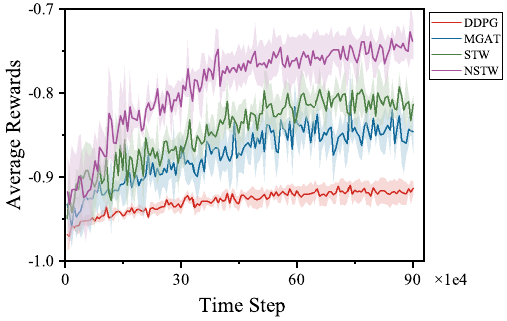}}
\caption{The average rewards  of different algorithms.}
\label{fig4}
\end{figure}

\begin{figure*}[bt!]
    \centering
    \includegraphics[width=1\linewidth]{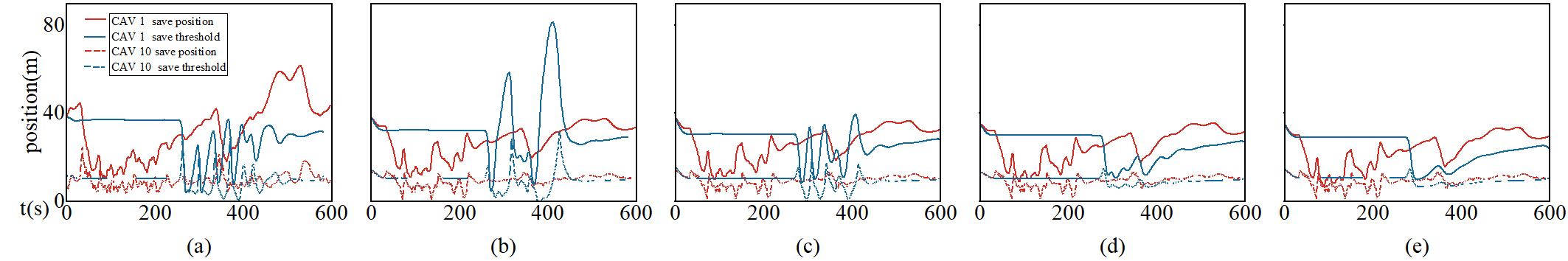}
    \caption{ Spatial distance between the lead and tail CAVs and their respective preceding vehicles under different algorithms: (a) IDM, (b) DDPG, (c) MGAT, (d) STW, (e) NSTW. }
    \label{fig:juli}
\end{figure*}
\begin{figure}[bt!]
    \centering
    \includegraphics[width=1\linewidth]{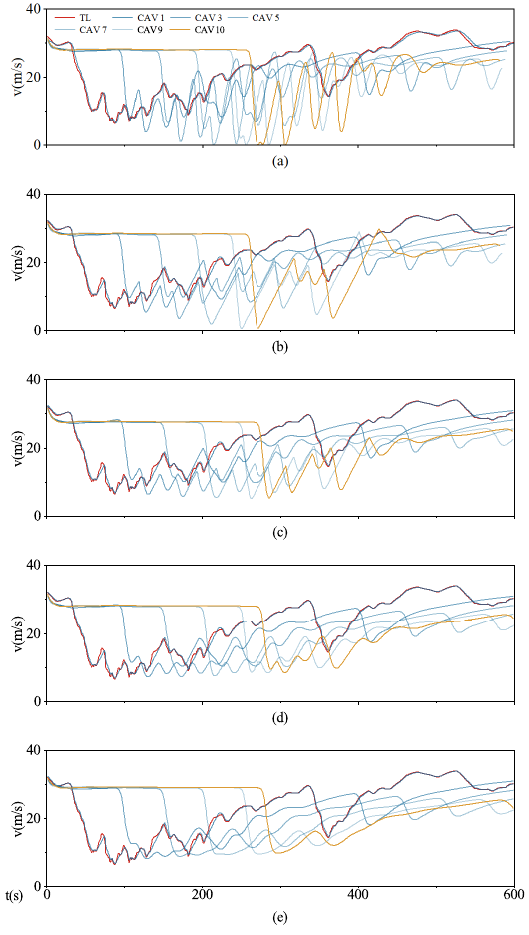}
    \caption{ Speed fluctuations under different algorithms: (a) IDM, (b) DDPG, (c) MGAT, (d) STW, (e) NSTW.}
    \label{fig:speed}
\end{figure}
\subsection{Evaluation parameters}
For the evaluation of the test results, a comprehensive suite of metrics has been adopted to thoroughly assess the effects of various methodologies on the performance of hybrid vehicle platoons:
\begin{enumerate}[(a)]
\item Spatial Distance Stability: Utilizing the safety reward function detailed in Section \ref{safe}, this metric evaluates the spatial distances between CAVs and the vehicle ahead at each moment, relative to the minimum safety distance. Consistent spatial distances not only demonstrate effective spacing control but also indicate enhanced safety levels.
\item Velocity Fluctuation Smoothness: This metric assesses traffic flow stability by analyzing the velocity fluctuations between the TL and the CAVs. Reduced velocity fluctuations suggest smoother vehicle platoon operations and, consequently, a more stable traffic flow.
\item Traffic Throughput: As defined by the task reward function in Section \ref{task}, traffic throughput is measured by plotting a spatiotemporal graph that maps the positions and times of all vehicles and calculating the total number of vehicles passing a specific location per hour. A higher throughput denotes greater traffic efficiency under the influence of the implemented algorithm.
\item Acceleration and Jerk Distribution: Based on the comfort reward function in Section \ref{confort}, this metric examines the distribution of acceleration and jerk across all vehicles. A focused distribution suggests that the algorithm enhances passenger comfort by minimizing abrupt vehicle movements.
\item Average Energy Consumption: According to the energy-saving reward function outlined in Section \ref{energy}, this measure calculates the total energy consumed by all vehicles—including driving, communication, migration, and computational energies (the latter three pertaining solely to CAVs). Vehicles are grouped into ten platoons led by CAVs, and the total energy consumption of each group is evaluated against its total distance traveled to derive the average energy consumption per unit distance. This metric serves to assess the energy efficiency of the decision-making strategies, with lower values indicating superior energy conservation.
\end{enumerate}

\section{Experimental Results}

\subsection{Comparative Algorithm Experiments}

\subsubsection{Training results}
As illustrated in Fig.~\ref{fig4}, the NSTW algorithm demonstrated superior performance, achieving the highest rewards with faster convergence and more stable training outcomes. This early advantage indicates the algorithm's capability to quickly adapt to the environment and optimize the driving strategy effectively. The NSTW algorithm not only ramped up faster but also exhibited lower standard deviation levels, underscoring its robustness.

\subsubsection{Testing results}
Fig.~\ref{fig:juli} illustrates that the IDM algorithm exhibits poor performance regarding fluctuation in following distances, negatively impacting safety. Conversely, the DDPG algorithm, although conservative, shows greater stability. The MGAT algorithm enhances fluctuation frequency but lacks efficiency in managing dynamic interactions between vehicles. Meanwhile, the STW algorithm significantly improves control over following distances, enabling more accurate adaptation to traffic fluctuations. Ultimately, the NSTW algorithm demonstrates the smoothest following distance curves, substantially enhancing driving smoothness and safety.
\begin{figure*}[bt!]
    \centering
    \includegraphics[width=1\linewidth]{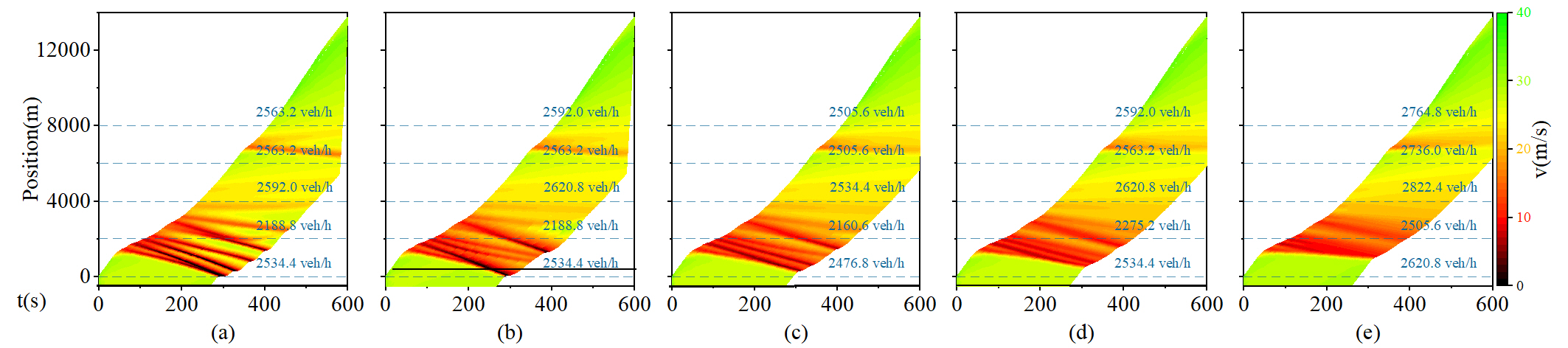}
    \caption{ Spatio-temporal heat maps showing traffic flow patterns: (a) IDM, (b) DDPG, (c) MGAT, (d) STW, (e) NSTW. Traffic waves are represented in red and black, with the bright green color representing the free flow, in addition to the blue dashed line showing the throughput of the traffic flow at a particular location.}
    \label{fig:heat}
\end{figure*}
\begin{figure*}[bt!]
    \centering
    \includegraphics[width=1\linewidth]{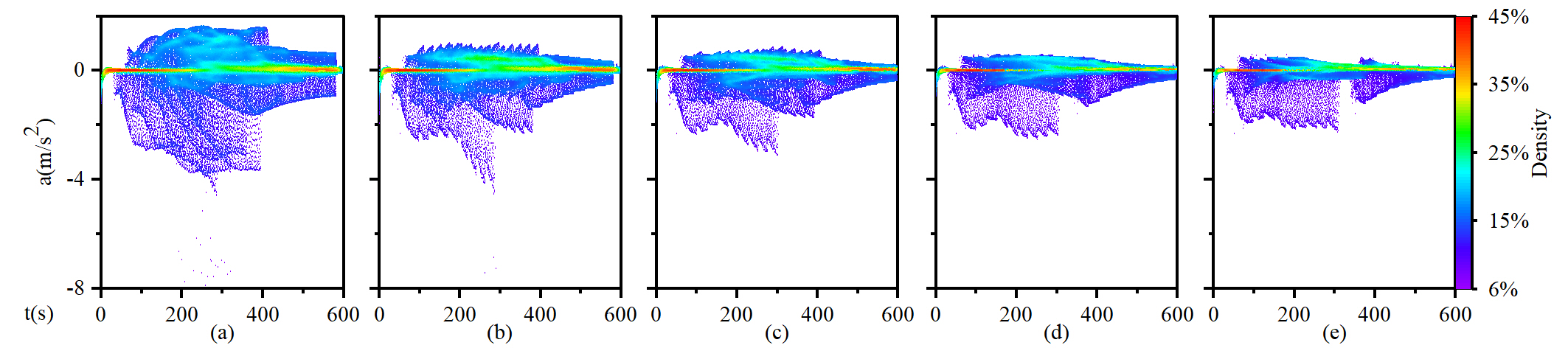}
    \caption{ Acceleration distributions highlighting the stability of  algorithms: (a) IDM, (b) DDPG, (c) MGAT, (d) STW, (e) NSTW. The colors represent the density of the current value in all current time data points, and the vertical spread of the colors indicates the range of variation in acceleration.}
    \label{fig:jia}
\end{figure*}
\begin{figure*}[bt!]
    \centering
    \includegraphics[width=1\linewidth]{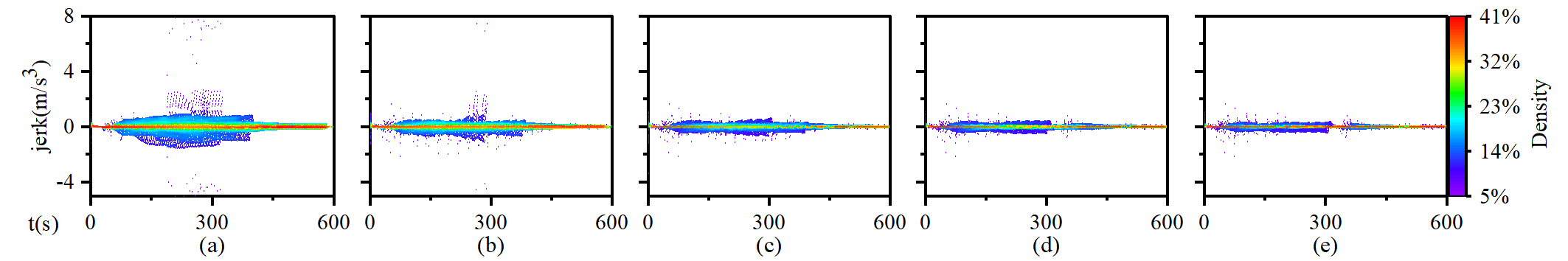}
    \caption{ Jerk distributions highlighting the stability of  algorithms: (a) IDM, (b) DDPG, (c) MGAT, (d) STW, (e) NSTW. }
    \label{fig:congjidu}
\end{figure*}
Fig.~\ref{fig:speed} illustrates the speed fluctuations of the TL and the following CAVs under various control strategies. The NSTW algorithm exhibits superior performance in stabilizing vehicle travel speeds compared to other strategies. It effectively minimizes vehicular interactions and oscillations within the traffic flow, thereby enhancing the overall efficiency and safety.

In Fig.~\ref{fig:heat}, the trajectories of individual vehicles are depicted as colored lines over time and space, where the color indicates the vehicle's speed. Typically, a "go-stop" wave pattern is observable in traffic flows. However, with the implementation of the NSTW algorithm, such patterns are virtually absent, resulting in a 14\% increase in throughput—significantly surpassing the performance of other algorithms in this domain and markedly enhancing capacity at virtual bottlenecks.

Fig.~\ref{fig:jia} shows that the acceleration distribution under the NSTW algorithm is highly concentrated, indicating that most vehicles maintain nearly uniform speeds with minimal acceleration or deceleration. This stable driving behavior not only improves ride comfort but also diminishes traffic volatility, which is crucial for preventing traffic congestion and enhancing the overall fluidity of traffic flow.

As depicted in Fig.~\ref{fig:congjidu}, the distribution characteristics of jerk, similar to acceleration, show a pronounced concentration around 0 m/s³, indicating a peak where most vehicles exhibit a preference for smooth driving to avoid substantial jerks. In Fig.~\ref{fig:congjidu}(e), the NSTW algorithm optimizes this distribution, with nearly all vehicles maintaining minimal variations in acceleration. This reflects a stable driving state that significantly aids in mitigating traffic congestion.

\begin{figure}[!t]
\centerline{\includegraphics[width=\columnwidth]{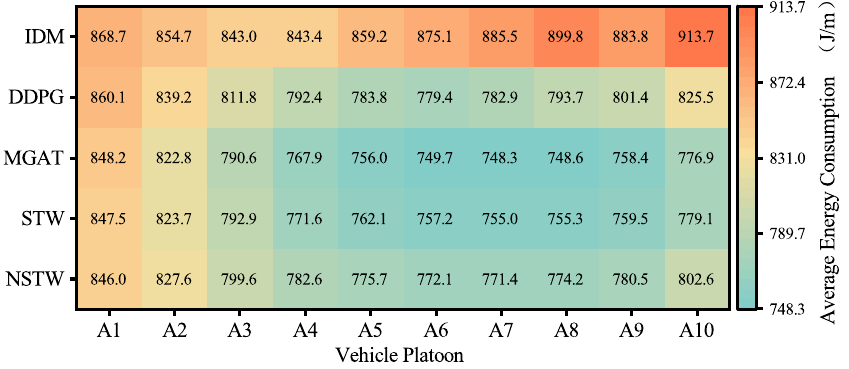}}
\caption{Average energy consumption diagram of 10 vehicle formations trained by 5 algorithms, with blue color representing smaller values of average energy consumption and red color representing larger values of average energy consumption.}
\label{fig8}
\end{figure}

As demonstrated in Fig.~\ref{fig:speed} and Fig.~\ref{fig8}, the NSTW algorithm mitigates unnecessary accelerations and decelerations, leading to a remarkably even distribution of average energy consumption. The maximum energy conservation reaches 125.6 J/m, cutting the average energy consumption by 9\% and increasing throughput by 10\% compared to the IDM. When juxtaposed with the MGAT and STW algorithms, the NSTW algorithm shows a modest increase in average energy consumption by only 1.6\%, yet it enhances throughput by 6\% to 16\%. These features render the NSTW algorithm more robust and stable amidst variations in traffic flow.

\begin{table*}
\caption{Testing evaluation results.\label{tab:table2}}
\newcolumntype{C}{>{\centering\arraybackslash}X}
\centering
\begin{tabular}{ccccccccccccccc} 
\hline
\centering
\multirow{2}{*}{\textbf{Algorithms}} & \multicolumn{3}{c}{$x (\rm{m}) $} &  $\overline q (\rm{veh/h})$ &\({v_{\min }}(\rm{m/{s}})\)
 &\multicolumn{2}{c}{${a \rm{(m/{s^2})}}$} & \multicolumn{2}{c}{$j \rm{(m/{s^3}) }$}  & \multicolumn{2}{c}{ ${E (\rm{J}}/{\rm{m}})$} & \multicolumn{3}{c}{ ${\theta _{{\rm{safe}}}} (\rm{m})$}\\
               &$\rm{max}$       & $\mu $   &$ \sigma$            & $\mu $ &$\rm{min}$                  &$\rm{max}$  &  $\rm{min}$                  & $\rm{max}$  & $\mu $                &$\rm{max}$  &  $\rm{min}$  &$\rm{max}$& $\mu $ &$\sigma $\\ 
\hline
\specialrule{0em}{2pt}{1pt}
     IDM								       &   61.76       &   9.57        & 29.80         &2505.6  &                0  & 1.56 &  -3.74               & 0.79 &0.23      &913.7 &843.0 &29.69 & 9.50 &3.43                \\    
\specialrule{0em}{1pt}{0pt}
     DDPG								      &   81.44         &  9.07        & 29.04       &2534.4  &                1.01   & 0.93 &  -2.47                & 0.51 &0.12      & 960.1 &779.4  &35.52  & 9.51 &3.42             \\

\specialrule{0em}{1pt}{0pt}
                 MGAT					&   39.40 	   &          \textbf{6.23}      &25.97  & 2449.0 &                	5.11   &0.74  &               	-2.36    & 0.50 &  0.09 & 848.2 &  \textbf{748.3} &18.79  & \textbf{9.44} &  2.24                 \\

\specialrule{0em}{1pt}{0pt}
                  STW					 &   35.56	   &          6.47       &24.55 & 2505.6 &                	\textbf{6.50}   &0.52  &               	-2.38    & 0.40 &  0.07 & 847.5 &  755.0  &14.78 & 9.49 &  \textbf{1.67  }              \\

\specialrule{0em}{1pt}{0pt}
                  NSTW				&   \textbf{34.75}	   &          6.64       &\textbf{22.92}   & \textbf{2678.4} &                	\textbf{6.50}   &\textbf{0.46}  &               	\textbf{-2.04}    & \textbf{0.30} &  \textbf{0.05 }& \textbf{846.0} &  771.4   &\textbf{14.76}      & 9.60 &  \textbf{1.67 }          \\


\specialrule{0em}{1pt}{1pt}
\hline
\end{tabular}
\begin{tablenotes}
\item\( x (\text{m}) \) is the distance to the preceding platoon for each platoon. \( \overline{q} (\text{veh/h}) \) is the traffic throughput. \( \theta_{\text{safe}} (\text{m}) \) is the safety threshold for CAVs.

\end{tablenotes}
\end{table*}
Table \ref{tab:table2} demonstrates that in mixed platooning environments, the NSTW algorithm achieves optimal performance across seven evaluative metrics. This superior performance suggests that the proposed framework effectively integrates nested graph representations, nested graph reinforcement learning, and spatiotemporal weighted multi-head graph attention mechanisms. These integrations considerably improve the decision-making robustness and overall effectiveness of platooning strategies.
\subsubsection{Energy consumption analysis}
\begin{figure}[!t]
\centerline{\includegraphics[width=\columnwidth]{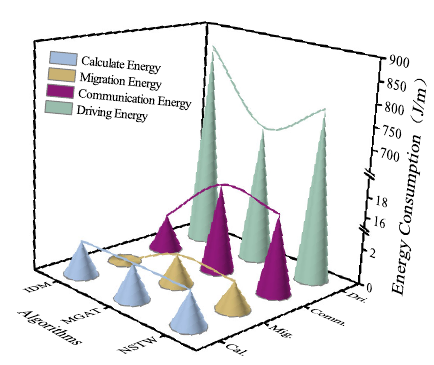}}
\caption{Schematic diagram of different energy consumption types.}
\label{fig:esort}
\end{figure}
Fig.~\ref{fig:esort} illustrates that both the MGAT and NSTW algorithms, while slightly increasing the energy consumed by communication, migration, and computation, significantly reduce the driving energy consumption compared to the IDM algorithm, thereby demonstrating substantial advantages. The test results suggest that the decision strategies proposed in this paper not only compensate for the energy expenditures of sensors, communication devices, and computing equipment but also offset the additional driving energy costs induced by the increased total vehicle mass. These strategies leverage a "less is more" approach to energy efficiency, markedly reducing overall energy consumption. Specifically, the NSTW algorithm reduces driving energy consumption by 80.4 J/m while adding only 16.6 J/m in additional energy costs, resulting in a net average energy savings of 63.8 J/m, thereby achieving true optimization of overall energy efficiency.
\begin{figure}[bt!]
    \centering
    \includegraphics[width=1\linewidth]{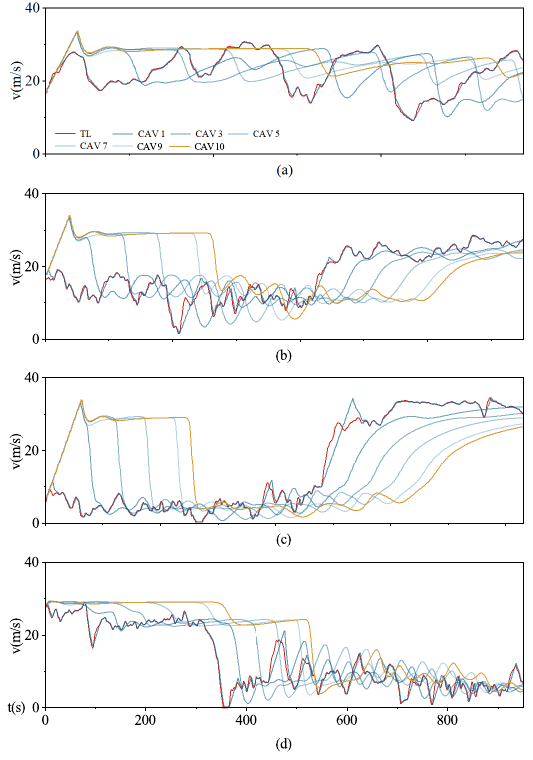}
    \caption{  Speed fluctuations under different algorithms: (a)  High-speed, (b) Low-speed, (c) Rapid acceleration, (d) Emergency braking.}
    \label{fig:sudu5}
\end{figure}
\begin{figure*}[bt!]
    \centering
    \includegraphics[width=1\linewidth]{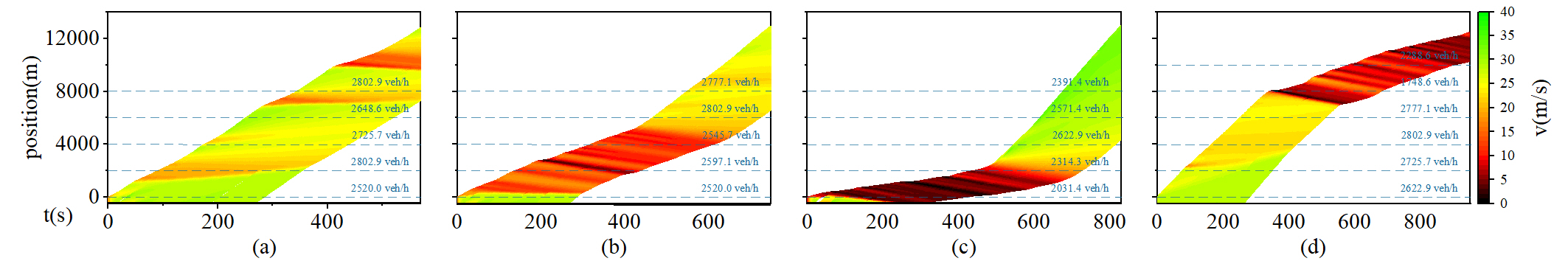}
    \caption{ Spatio-temporal heat maps showing traffic flow patterns: (a)  High-speed, (b) Low-speed, (c) Rapid acceleration, (d) Emergency braking. }
    \label{fig:st5}
\end{figure*}
\begin{figure}[bt!]
    \centering
    \includegraphics[width=1\linewidth]{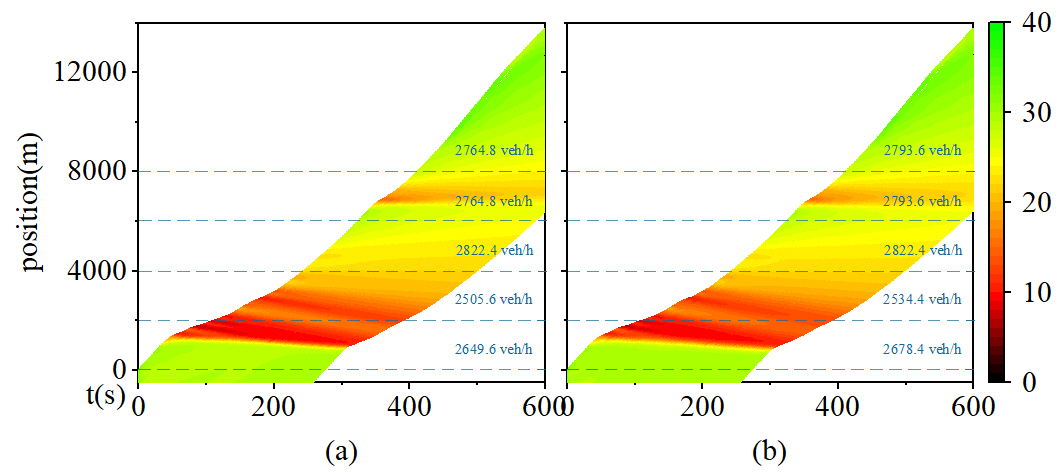}
    \caption{ Spatio-temporal heat maps showing traffic flow patterns: (a) 10\% penetration rate, (b) 20\% penetration rate. }
    \label{fig:st6}
\end{figure}
\begin{figure}[bt!]
    \centering
    \includegraphics[width=1\linewidth]{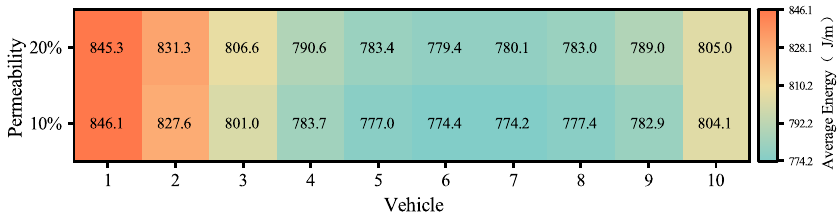}
    \caption{ Average energy consumption diagram of different penetration rate.}
    \label{fig:sten6}
\end{figure}
\subsection{Generalizability Testing Experiments}
As illustrated in Figures \ref{fig:sudu5} and \ref{fig:st5}, the NSTW algorithm exhibits exceptional adaptability and stability across a variety of driving scenarios. Under both high-speed and low-speed conditions, the algorithm effectively manages the velocity fluctuations of CAVs, maintaining consistently smooth speed profiles. This demonstrates the algorithm’s robust foundational adaptability to dynamic conditions. Notably, in scenarios involving rapid acceleration and deceleration, although significant speed changes occur, the transitions are managed smoothly, significantly mitigating potential safety risks associated with sudden velocity changes.

The spatiotemporal heatmap in Fig.~\ref{fig:st5} further substantiates the algorithm’s efficacy under varying traffic flow conditions. In high-speed environments, the algorithm swiftly adapts, maintaining high vehicle velocities, thus ensuring a fluid traffic flow in high-throughput situations. Under constrained low-speed conditions, it displays considerable generalization capabilities, effectively smoothing out "stop-and-go" fluctuations and reducing traffic congestion.

Even in extreme acceleration and deceleration conditions, where traffic variability increases, the NSTW algorithm continues to maintain fundamental traffic flow stability. With a minimum throughput sustained at 1748.6 vehicles per hour, the algorithm successfully prevents large-scale stopping incidents, markedly improving the system’s safety and efficiency.

These experimental outcomes not only demonstrate the NSTW algorithm’s superior performance under diverse dynamic traffic conditions but also emphasize its potent generalization capability, even without specific scenario training. This capability is vital for navigating the continuously evolving traffic conditions in real-world environments, aiding autonomous driving systems in maintaining operational efficiency and safety in complex and variable road conditions.
\subsection{Permeability Ablation Experiments}

Fig.~\ref{fig:st6} and Fig.~\ref{fig:sten6} provide an in-depth analysis of the impact of varying CAVs penetration rates on traffic flow efficiency and energy efficiency by comparing vehicle spatiotemporal distributions and energy consumption across different scenarios. Fig.~\ref{fig:st6} illustrates that a higher penetration rate of CAVs significantly enhances traffic throughput, demonstrating improved efficiency through more effective road resource utilization. Particularly in Fig.~\ref{fig:st6}(b), a 20\% CAVs penetration rate not only increases the duration and spatial extent of high-flow states but also showcases the potential of CAVs in managing complex traffic scenarios.

Fig.~ \ref{fig:sten6} compares vehicle platoon energy consumption at 10\% and 20\% CAVs penetration rates, revealing an intriguing trend. Theoretically, a higher CAVs penetration should lower energy consumption by promoting smoother driving and more uniform acceleration. However, a slight increase in energy consumption is observed, suggesting that despite the substantial improvements in traffic flow stability and vehicular coordination provided by CAVs, the additional energy expenditures for driving, communication, and computation may still negate the potential energy savings. This finding emphasizes the importance of comprehensive system-wide energy consumption assessments in the design and implementation of CAV systems to ensure the optimization of energy efficiency.

These analyses highlight that while promoting CAVs to boost traffic efficiency, it is crucial to consider the multidimensional aspects of energy consumption, particularly in environments with high CAV penetration. Further research should focus on how to achieve true energy efficiency maximization through technological innovations and system design optimization while enhancing traffic flow efficiency.
\section{Conclusion}
To tackle the collaborative control challenges posed by vehicular heterogeneity in large-scale mixed platooning, we propose a multi-objective optimization decision strategy rooted in nested graph reinforcement learning. This strategy enables platoons to dynamically optimize their decision-making processes, ensuring efficient energy use while minimizing traffic congestion. Our simulation experiments provide a thorough evaluation of key performance indicators such as traffic flow, vehicle speed distribution, energy consumption efficiency, and congestion scenarios. 
\begin{enumerate}
\item Compared to the IDM algorithm, the NSTW algorithm significantly reduces the average spatial distance between vehicles by 30\%, lowers the standard deviation of acceleration by 63.77\%, enhances the overall traffic throughput by 10\%, and reduces the average energy consumption by 9\%.
\item The proposed NSTW algorithm decreases driving energy consumption by 80.4 J/m while only adding an additional energy cost of 16.6 J/m. Moreover, increasing the penetration rate of CAVs can significantly boost traffic throughput, but it may also increase energy consumption, necessitating comprehensive optimization to achieve true energy efficiency maximization.
\item The NSTW algorithm exhibits exceptional generalizability and stability across various driving scenarios, maintaining traffic flow stability even under extreme acceleration and deceleration conditions, thus significantly enhancing the system's safety and efficiency.
\end{enumerate}

In future research, we plan to address the observed increase in energy consumption by investigating different communication and computation strategies to find an optimal balance between energy efficiency and system performance. Additionally, considering the diversity of urban environments and traffic densities, we will also validate and refine the model's adaptability and accuracy in various practical application scenarios. 
Ultimately, by collaborating with industry partners, we aim to commercialize our research findings, providing innovative support for urban traffic management systems.

\bibliographystyle{IEEEtran}

\end{document}